\documentclass[%
reprint,
amsmath,amssymb,
aps,showkeys
]{revtex4-2}

\usepackage{graphicx}% Include figure files
\usepackage{dcolumn}% Align table columns on decimal point
\usepackage{bm}% bold math
\usepackage{hyperref}% add hypertext capabilities
\hypersetup{colorlinks,linkcolor={blue},citecolor={red},urlcolor={blue}}
\usepackage[mathlines]{lineno}% Enable numbering of text and display math
\begin{document}

\preprint{APS/123-QED}

\title{Dynamics of Brownian particles in asymmetric confinement: Insights into Entropic Stochastic Resonance}

\author{Syed Yunus Ali}
\email{CY18D503@iitp.ac.in}
\affiliation{Department of Chemistry and The Center for Atomic, Molecular, and Optical Sciences \& Technologies, Indian Institute of Technology Tirupati, Yerpedu 517619, Andhra Pradesh, India.}

\altaffiliation[Currently at]{ Department of Physics, Indian Institute of Technology Bombay, Mumbai 400076, Maharashtra, India.}

\date{\today}

\begin{abstract}
We explore the effect of asymmetry in the thermodynamic response (work done) of an overdamped Brownian system driven by a time-periodic field when the particle is confined inside a bilobal irregular structure. The spatial irregularity of the asymmetric confinement results in an effective asymmetric entropic bistable potential along the direction of transport. We investigate how the frequency of the periodic field and the intensity of the noise impact the average work done, focusing on its potential as a key metric for examining Entropic Stochastic Resonance (ESR). The study highlights the impact of confinement asymmetry on reducing the average work done. Furthermore, we observe a transition of average work done from a state dominated by energy to one dominated by entropy, as we manipulate the magnitude of the transverse force. In addition, an alternative quantity called the mean free flight time ($T_{MFFT}$) is proposed to describe the ESR in the presence of asymmetry.
\end{abstract}

%Use showkeys class option if keyword
                              %display desired
\maketitle

\section{\label{sec:level1}Introduction}
A Brownian particle may exhibit an enhanced response to periodic forcing, leading to an amplification of the input at an optimal noise level. This resonant phenomenon observed in non-linear systems that are regularly modified, where noise can amplify input at an appropriate noise level, is known as Stochastic Resonance (SR) \cite{Benzisiam1983,gammaitonirevmodphys1998, Wellens2003iop}. During the last two decades, SR-like phenomena have received considerable attention in a wide range of disciplines, including physics, chemistry, engineering, and biological research. Mainly, SR involves an energetic potential system \cite{hanggi2002chemphyschem, Yasuda2008prl,badzey2005nat, Vilar1996prl,wu2008physletta}. However,  for many practical systems like biological systems, soft condensed matter system entropic barriers \cite{zwanzig_1992_diffusion,reguera_2001_kinetic,kalinay_2005_extended,kalinay_2006_corrections,ai_2006_current,reguera_2006_entropic,burada_2007_biased,burada_2008_entropic_prl,burada_2009_double,burada2009diffusionchemphyschem,mondal_2010_diffusion,mondal_2010_entropic_noise,das_2012_shape,das_2012_hyst,das_2012_logic,kalinay2006pre,zhou_2010_diffusioninfluenced,bnichou_2010_geometrycontrolled,mondal_2011_asymmetric,mondal_2016_ratchet,Carusela2017jcp,Das2015pre,andrsarangorestrepo_2020_entropic,Rubi_2019epl} play an essential role in many processes like the flow of nutrients into or out of the bloodstream, the transport of ions or macromolecules across membranes, and signal transduction in the synapses \cite{zhou2010jpcl,licata_2016_diffusion,pagliara_2014_diffusion,arangorestrepo_2021_enhancing,santamaraholek_2019_review,radi_1998_peroxynitrite,ricciardi_2013_diffusion,kleinfeld_1998_transport,dagdug2022blocker}. These entropic obstacles unveils entropy-driven stochastic phenomena, including entropic analogs of stochastic resonance \cite{burada_2008_entropic_prl,burada_2009_double}, resonant activation \cite{mondal_2010_entropic_resonant}, asymmetric stochastic localization \cite{mondal_2011_asymmetric}, dynamic hysteresis \cite{das_2012_hyst,mondal_2012_entropic}, entropic ratchet effect \cite{marquet2002,kalinay2023pre,kalinay2022pre,kalinay2021pre,kalinay2018pre}, particle mobility \cite{ralf2002prl}, and noise-induced nonequilibrium transitions \cite{mondal_2010_entropic_noise}, rectification \cite{malagaretti2012pre,malgaretti2013jcp,Dadug2012jcp}, entropic entrapment \cite{martens2013prl,ai2013jcp,shi2014pre}, and particle separation \cite{regura2012prl,motz2014jcp}.\\

The phenomenon of entropic stochastic resonance (ESR) was initially reported by Burada et al. (2008) in their study \cite{burada_2008_entropic_prl}. ESR emerges when the Brownian particles are exposed to a periodic force in the horizontal direction within a confined dumbbell-shaped geometry. It is distinguished by the appearance of a single peak in the spectral amplification at the appropriate noise strength values. Brownian particles driven by a constant bias can exhibit double ESR \cite{burada_2009_double}, while Brownian particles confined to two distinct regions divided by a porous membrane can show a type of ESR dependent on a geometric effect. Recent research has demonstrated that Brownian particles confined in a periodic channel can be trapped by ESR \cite{shi2014pre}. A direct demonstration of the ESR mechanism is provided \cite{Zhu2022prl}. However, for the Brownian particle to capture the entropic barrier, it needs constant bias force to act in the vertical direction. Effective entropic potential (ESR) arises as particles move in confined regions. The assumption of local equilibrium in the transverse direction can reduce two-dimensional or three-dimensional dynamics to one-dimensional dynamics. However, investigations on the impact of time-periodic forces on entropic transport continue to be conducted \cite{he2010pre,wu2015physletta, Ye2019physscrp}. While research on ESR demonstrates that symmetry breakdown generated by the longitudinal or transverse biasing force is required for ESR \cite{burada_2008_entropic_prl,burada_2009_double,shi2014pre}, the orthogonal periodic driving force can improve longitudinal entropy transmission \cite{he2010pre}. Researchers have conducted studies on the dependence of entropic stochastic resonance (ESR) on various factors, such as the shape of boundaries  \cite{burada_2008_entropic_prl,ghosh_2010_geometric,mei2021rstc}, external forcing \cite{burada_2009_double,ding2015jcp,du2021rstc}, and properties of noise \cite{liang2010cpl,xu2020cjp}. Additionally, researchers have explored the potential application of entropic stochastic resonance (ESR) in particle manipulation \cite{shi2014pre}.\\
\par
Gammaitoni et al. showed that the residence time distribution $N(T)$ exhibits resonant behavior when the driving frequency changes \cite{gammaitonirevmodphys1998}. 
Marchesoni et al. demonstrated through numerical analysis that the Schmitt trigger's $N(T)$ displays a frequency peak in both weakly and strongly driven systems \cite{Marchensoni2009pre}. The association between Stochastic Resonance (SR) and the synchronization of passages between different wells can also be described by the hysteresis loop area \cite{mahato1994pre, mahato1997pre, mahato1997modphyslettb, mahato1998physa}. The magnitude of this loss can be regarded as an indicator of SR, known as a bonafide resonance. The input energy is also a suitable metric of SR \cite{Iwaiphysa2001, Iwaijpsj2001, saikia_2007_work,sahoo_2008_stochastic, jop2008work, Ali2024jpcb}. This energy is equal to the work done by the external agent that drives the potential periodically. The input energy not only exhibits peaking behavior with noise intensity, but also exclusively considers interwell behavior. In the conventional SNR, both intrawell as well as interwell motion are taken into account. Hence, for the minor driving frequency and noise strength, the motion is predominated by the interwell oscillations \cite{Iwaiphysa2001, Iwaijpsj2001}. So, the peak in the input energy is a strong indicator of the match between the escape rate and the external driving frequency. The input energy distribution, with a typical most enormous shoulder at stochastic resonance, provides valuable information on stochastic resonance behavior. Recently, we also discovered that the input energy, that is, the work done and the heat absorbed over a cycle, could also act as a quantifier of the ESR \cite{Ali2024jpcb}. where we described the stochastic resonance and bonafide resonance phenomena by the mean value (as well as fluctuation) of work done and absorbed heat over a period as a function of noise strength and input frequency.\\

In this paper, we show how the asymmetry in the structure affects the extracted work. In irregular asymmetric bilobal, how does the ESR weaken with increasing extent of asymmetry? Our paper is organized as shown below. In Sect.~\ref{sec:level2}, we review the theoretical description for the overdamped Brownian motion of a particle in a two-dimensional symmetric bilobal confined geometry with an external periodic force, and we review how the effective potential varies as the asymmetry in the confinement increases. In Sec.~\ref{sec:level3}, we discussed our results. We conclude the paper in Sec.~\ref{sec:level4}.

\section{\label{sec:level2}The Model}

\begin{figure}[!h]
 \centering
\includegraphics[width=0.45\textwidth]{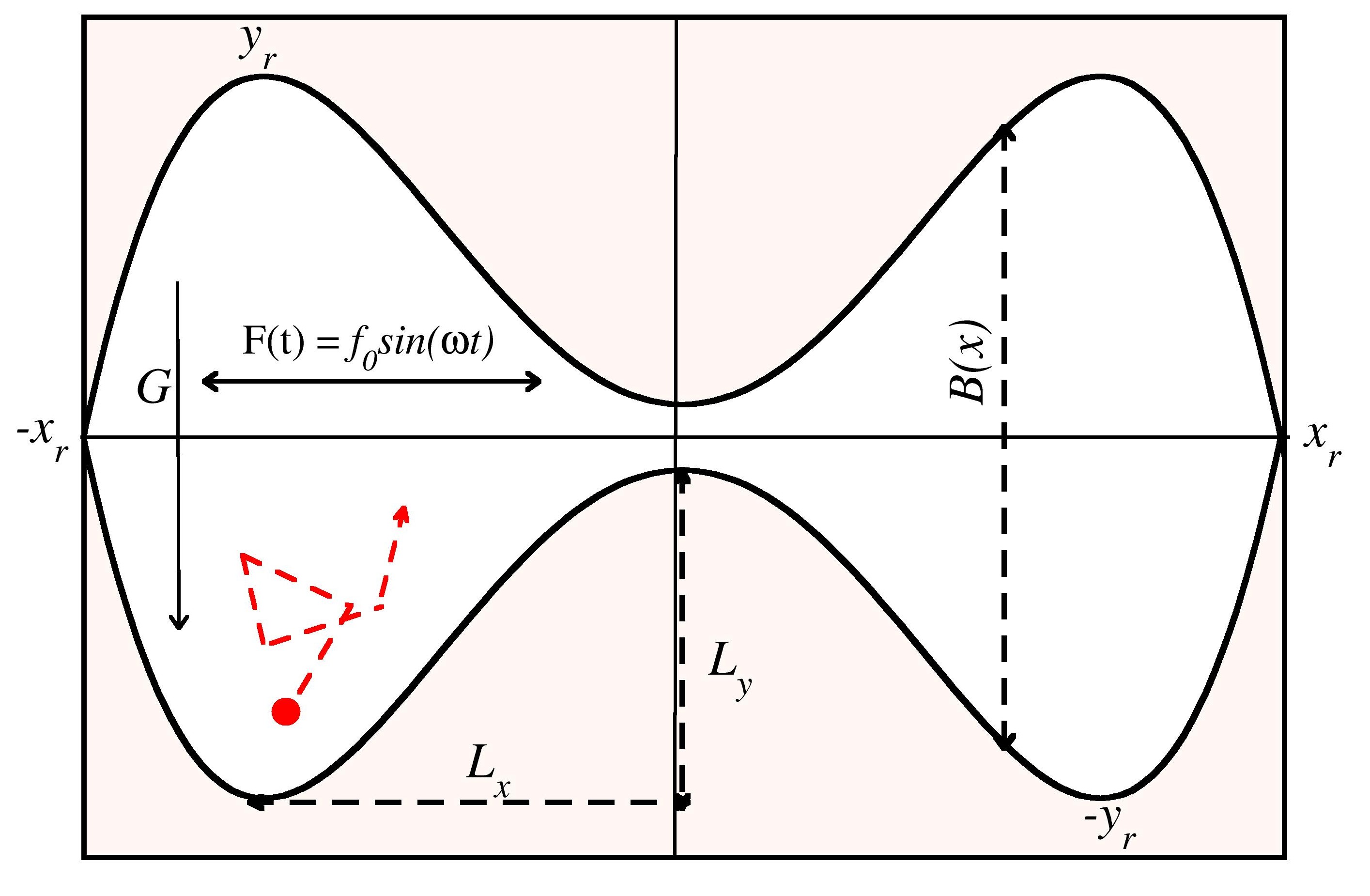}
\caption{Schematic diagram of a Brownian particle in 2-D bilobal confinement. $F(t) = (f_0\sin(\omega t))$ and $G$ represent the external periodic force in the direction ($x$) and the constant bias force in the transverse direction ($y$), respectively. $B(x)$ is the local half width at position $x$.}
\label{f1}
\end{figure}

As shown in Fig.~\ref{f1}, a Brownian particle is contained in a two-dimensional bilobal enclosure. The dynamics of the particle in an overdamped state can be represented by the Langevin equation provided below,
\begin{eqnarray}\label{e1}
\gamma (d\overrightarrow{r}/dt)=-\overrightarrow{e_{x}}F(t)+\sqrt{\gamma
k_{B}T} \overrightarrow{\xi}(t)-G\overrightarrow{e_{y}}\;,
\end{eqnarray}
where $\overrightarrow{r}$ represents the position vector of the particle. $\gamma$ denotes the friction coefficient; $\overrightarrow{e_{x}}$, $\overrightarrow{e_{y}}$ are the unit vectors along $x$ and $y$ directions, respectively. The particle experiences an additive thermal noise $\overrightarrow{\xi}$ and a periodic drive force $F(t)$ along the x-axis, and a constant transverse bias force $G$ along the $(y)$ direction. $\overrightarrow{\xi}$ is a Gaussian white noise with zero means and obeys the fluctuation-dissipation relation $\langle  \xi_{i}(t)\xi _{j}(t')\rangle=2\delta_{ij}\delta(t-t')$, for $i,j=x,y$. The external periodic driving force along the $x-$direction is denoted by $F(t) = f_0 \sin(\omega t)$, where $f_0$ and $\omega$ signify the amplitude and frequency of this external periodic driving force, respectively.
The imposition of reflecting boundaries, which have the form $B_{l}(x)=ax^4-2ax^2-c/2=-B_u$, can explain the confinement of a system. Where the aspect ratio $a= L_{y}/L_{x}$, $x$ and $y$ are position coordinates, and $B_{l}$, $B_{u}$ correspond to the lower and upper boundary functions of the system. $L_x$ is the distance between the position of the bottleneck and the position of the maximal width; $L_y$ is the narrowing of the boundary function; and $c$ is the width available at the bottleneck. Consequently, $B(x)=(B_{u}(x)-B_{l}(x))/2$ gives the local half-width of the structure. When the transverse force $(G)$ is minimal, an entropic potential controls the particle's behavior. In the limit of the large transverse force $(G)$, the particle's motion is restricted to the bottom boundary of the confining structure. Such overdamped motion of the particle can alternatively be described using the following 2-D Smoluchowski equation \cite{hrisken_1996_the,burada_2008_entropic_prl,burada2009diffusionchemphyschem}:
\begin{eqnarray}\label{e2}
\frac{\partial p(x, y, t)}{ \partial t} & = & D\frac{\partial}{\partial x} \left \{  e^{\frac{-\psi(x,y)}{D}} \frac{\partial}{\partial x} e^{\frac{\psi(x,y)}{D}}p(x,y,t)\right \} \\ \nonumber & + & D\frac{\partial}{\partial y}\left \{ e^{\frac{-\psi(x,y)}{D}} \frac{\partial}{\partial y} e^{\frac{\psi(x,y)}{D}}p(x,y,t)\right \},
\end{eqnarray} Owing to the Fick-Jacobs approximation, the 2-D description Eq.~\ref{e2} can be reduced to a one-dimensional effective Fokker-Planck equation in the absence of a temporal periodic forcing along the $x-$direction, i.e., $F(t) = 0$. \cite{zwanzig_1992_diffusion,reguera_2001_kinetic,kalinay_2005_extended,kalinay_2006_corrections,ai_2006_current,reguera_2006_entropic,burada_2007_biased,burada_2008_entropic_prl,burada_2009_double,burada2009diffusionchemphyschem,mondal_2010_diffusion,mondal_2010_entropic_noise,das_2012_shape,das_2012_hyst}

\begin{eqnarray}\label{e3}
\dfrac{\partial{P(x,t)}}{\partial t} =  \dfrac{\partial}{\partial
x}\left[D\dfrac{\partial}{\partial x}P(x,t)+A^{'}(x)P(x,t)\right].
\end{eqnarray}
Here $P(x,t)$ is the probability distribution function in reduced dimension, and the prime refers to the derivative with respect to $x$.
$A(x)$ is the effective entropic potential in reduced dimension \cite{reguera_2001_kinetic,kalinay_2005_extended,kalinay_2006_corrections,ai_2006_current,kalinay2006pre,reguera_2006_entropic,das_2012_shape,das_2012_logic,das_2012_hyst}, which has a bistable form for this bilobal confinement and reads as
\begin{eqnarray}\label{e4}
A(x) = -D\ln\left[\frac{2D}{G}\sinh \left(\frac{GB(x)}{D}\right)\right].
\end{eqnarray}
Depending on the transverse bias force $(G)$, thermal energy $(D)$, and local half-width of the system $(B(x)$, the ratio $\frac{GB(x)}{D}$ can be used to tune the appropriate regimes of the effective potential that controls the dynamics. When $\frac{GB(x)}{D}\gg 1$ Eq.~(\ref{e4}) yields $A(x)=-GB(x)$, it is the energy-dominated potential, while when $\frac{GB(x)}{D}\ll 1$ Eq.~(\ref{e4}) yields $A(x)=-D\ln(2B(x))$, it is independent of $G$. Therefore, the Brownian particle trapped in bilobal confinement experiences an effective geometric potential in reduced dimension as a function of the irregularity in the local half-width of the captivity. This potential resembles a bi-stable trap.
\par
\begin{figure}[!ht]
 \centering
\includegraphics[width=0.46\textwidth]{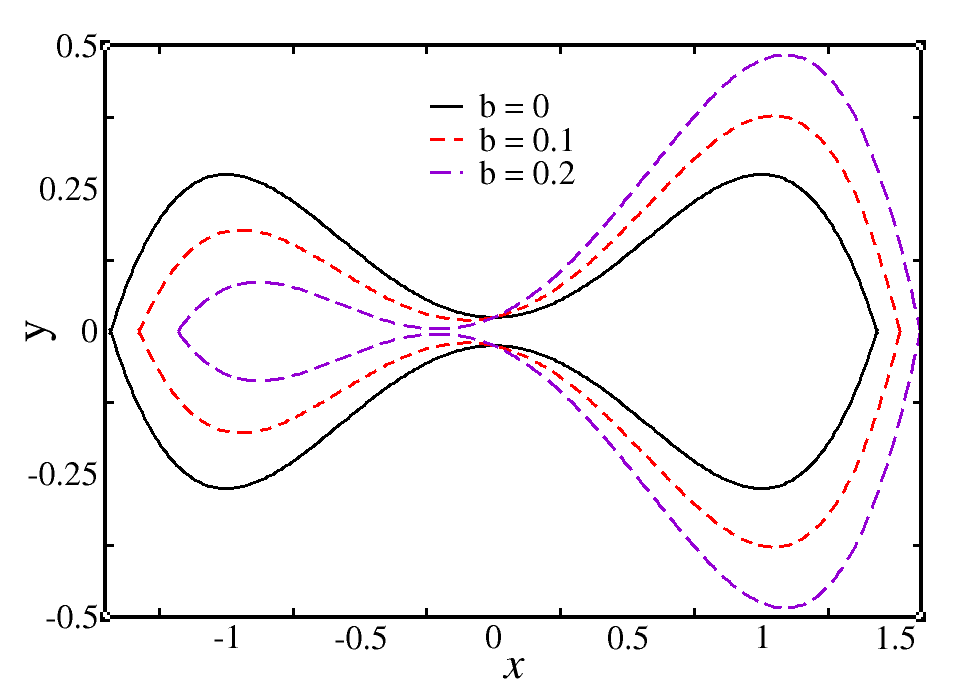}
\includegraphics[width=0.45\textwidth]{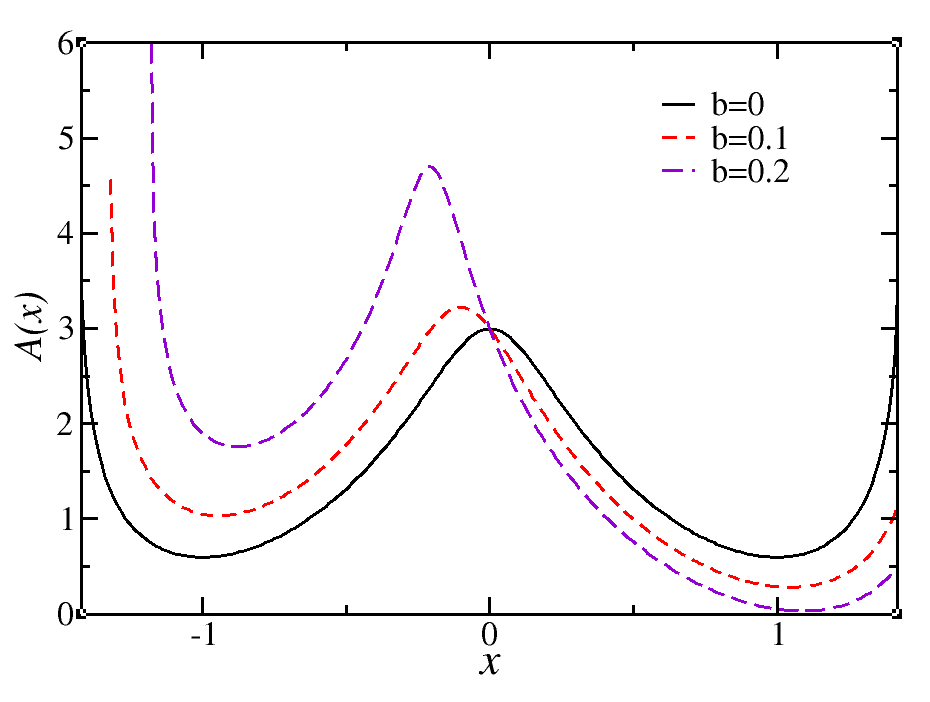}
\caption{Upper Panel:The schematic representation illustrates the bilobal confinement, with the symmetric bilobal confinement depicted by the solid black line. The confinement shape changes as the asymmetric parameter ($b$) is adjusted. Lower Panel: The effective potential undergoes alterations with variations in the asymmetric parameter ($b$). Parameter set chosen as $a = 0.25$, $c = 0.05$, $D = 1$, $\gamma = 1$.}
\label{f2}
\end{figure}

Next, we consider a scaled asymmetric bilobal confinement driven by the external field $F(t)$. We make the asymmetric bilobal confinement by incorporating  asymmetric reflecting boundaries, which have the form
\begin{equation}\label{e5}
    B_{l}(x) = ax^4-2ax^2-bx-c/2=-B_{u}(x)
\end{equation}
which can explain the confinement of a system. How the structure of the confinement changes with the change of asymmetric parameters is shown in Fig.~\ref{f2}.
We have shown that the average work done and absorbed heat in a periodically driven overdamped Brownian particle in a bilobal enclosure with uneven boundaries can be a good quantifier for studying entropic stochastic resonance \cite{Ali2024jpcb}. For our present purpose, we'll concentrate on the effect of asymmetric parameter $b$ on the work done over a cycle. The work done or the input energy ($\tau_{\omega}=\frac{2\pi}{\omega}$) to the system under the influence of the external drive is defined as \cite{sekimoto1997jpsj,sekimoto1999jpsj,dan2005bona,sahoo_2008_stochastic,saikia_2007_work}
\begin{eqnarray}\label{e6}
\begin{split}
W & = \int^{t_{0}+\tau_{\omega}}_{t_{0}} \frac{dU}{dt}dt^{'}\\
 & =-f_{0}\omega\int^{t_{0}+\tau_{\omega}}_{t_{0}}x(t^{'})\cos(\omega t^{'})dt^{'}
\end{split}
\end{eqnarray}
where $U(x,t)$ is the time-dependent potential $U(x,t)=A(x)-xF(t)$. It is necessary to evolve the system while disregarding any initial transients to calculate the value of work done $(W)$.

\section{Results and Discussion\label{sec:level3}}
\begin{figure}[!ht]
    \centering
    \includegraphics[width=0.5\textwidth]{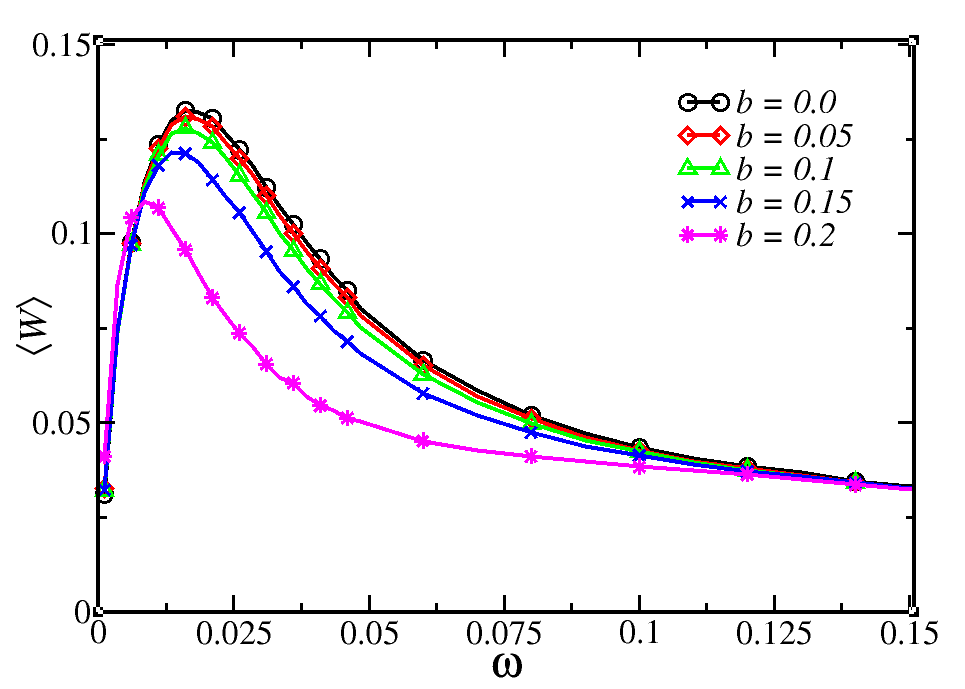}
    \caption{The figure of the variation of $\langle W \rangle$ with the frequency of driving $(\omega)$ for different value of asymmetric parameter, $b$. The parameter set chosen as $a = 0.25$, $c = 0.05$, $G = 0.0$, $f_0 = 0.05$, $d = 0.01$.}
    \label{f3}
\end{figure}

We calculate the average work done over the cycle numerically ($\langle W \rangle$) and vary with the frequency of the periodic force ($\omega$), the noise intensity $(D)$, and the transverse bias force $(G)$. Our numerical analysis starts by varying the average work ($\langle W \rangle$) with the frequency of the periodic force ($\omega$) for different asymmetric parameters $(b)$. The results are shown in Fig.~\ref{f3}.
In the limit of very high and low driving frequencies, the average work done is almost zero \cite{Ali2024jpcb}. For an intermediate frequency, the work done over a cycle assumes a nonzero value and exhibits a maximum \cite{sahoo_2008_stochastic,dan2005bona,jop2008work, Ali2024jpcb}. Thus $\langle W \rangle$  acts as quantifiers of bonafide stochastic resonance \cite{dan2005bona,sahoo_2008_stochastic,Ali2024jpcb}. An essential observation drawn from Fig.~\ref{f3} is the evident trend: as the degree of asymmetry represented by the parameter $'b'$ increases, there is a conspicuous decrease in the peak associated with the maximum work done. In other words, the distinctiveness of the bonafide stochastic resonance (SR) phenomenon becomes less prominent compared to the scenarios where $b = 0$ or is exceedingly low. Upon reaching a higher value of the asymmetry parameter $b$, the peak shifts to lower values of the periodic driving frequency. The observed weakening of synchronization is attributed to the inherent discrepancy induced by the nonzero asymmetry parameter. Specifically, this discrepancy manifests in the mean passage time from the right lobe to the left lobe differing from the mean passage time in the reverse direction—from the left lobe to the right lobe.

\begin{figure}[!ht]
    \centering
    \includegraphics[width=0.5\textwidth]{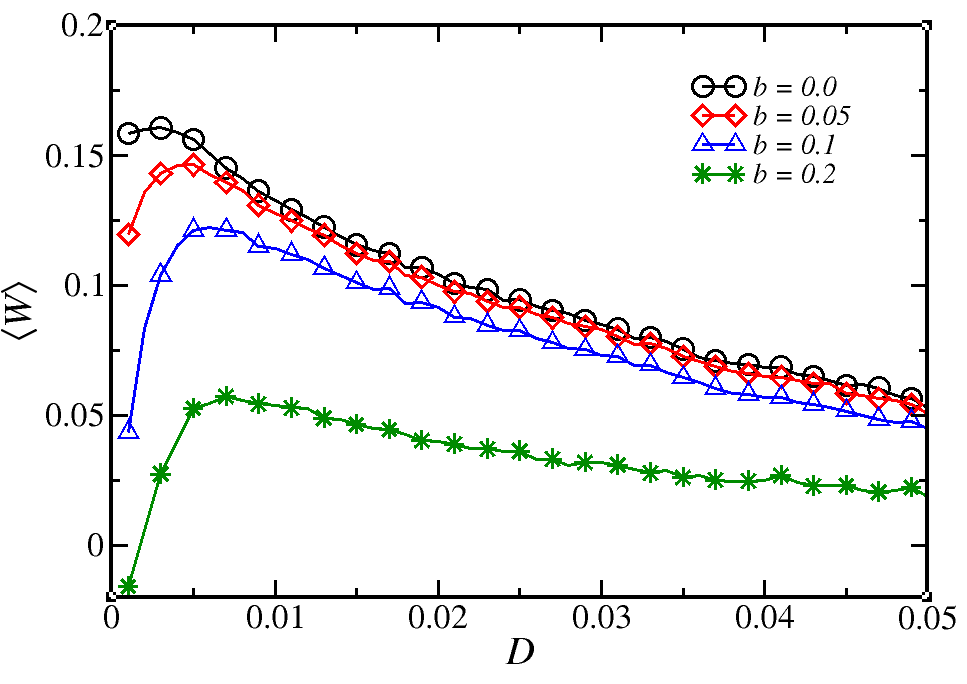}
    \caption{The variation of the average work done ($\langle W \rangle$) with the noise intensity ($D$) for different values of asymmetric parameter $b$. For all cases, $a = 0.25$, $c = 0.05$, $f_{0} = 0.05$, $G = 0.1$ and $\omega = 0.015$.}
    \label{f4}
\end{figure}

Moving forward, we find out the pivotal role thermal noise strength plays in producing such a nontrivial response. Our focus centers on scrutinizing the variations in the average work done, denoted as $\langle W \rangle$, as a function of the noise strength parameter ($D$), considering various values of the asymmetric parameter $(b)$. The outcomes of this investigation are presented in Fig.~\ref{f4}. In our analysis, holding the driving frequency constant and given a nonzero value of transverse bias force $(G)$, we observe that the work done exhibits a turnover passing through a maximum \cite{Ali2024jpcb}. Intriguingly, the degree of asymmetry $(b)$ exerts a noteworthy influence on the magnitude of work done $(\langle W \rangle)$. As the extent of asymmetry increases, there is a discernible reduction in the work done. This intriguing observation aligns with our earlier discussion regarding the mismatch in the transition time scales between the wells. The increase in asymmetry $(b)$ appears to hinder the efficiency of inter-well transitions, contributing to a reduction in the overall work done.
This observation supports the most efficient synchronized escape over the entropic potential well under optimal noise in the presence of a given strength and frequency of external driving. 

\begin{figure}[!ht]
    \centering
    \includegraphics[width=0.5\textwidth]{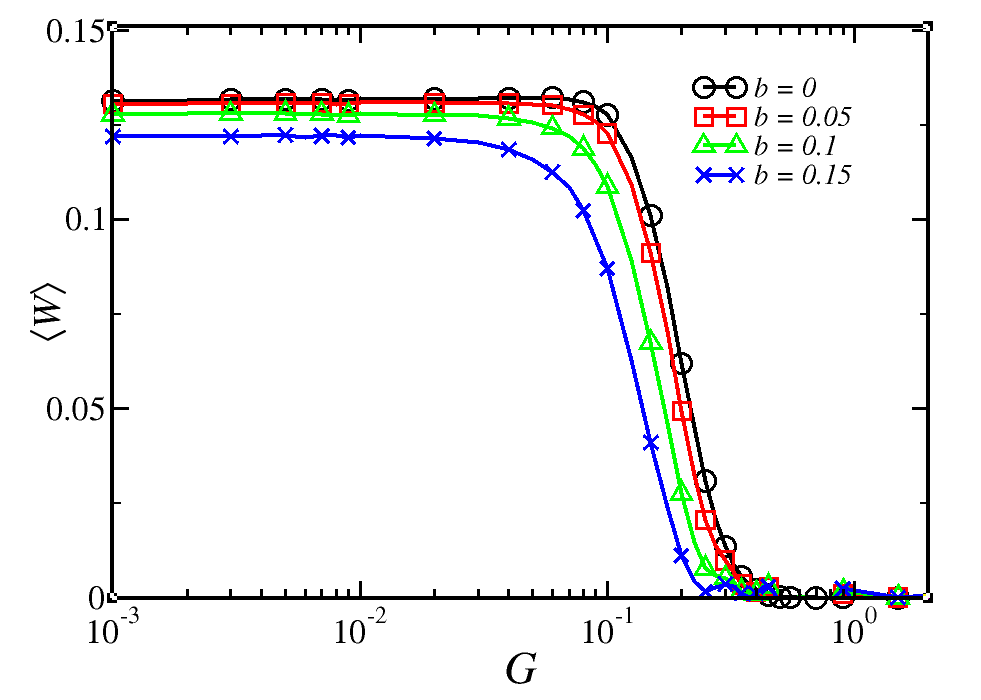}
    \caption{The variation of the average work done ($\langle W \rangle$) with the transverse force field ($G$) for different values of asymmetric parameter $b$. For all cases $a = 0.25$, $c = 0.05$, $f_{0} = 0.05$ and $\omega=0.03$.}
    \label{f5}
\end{figure}
In Fig.~\ref{f5}, we present a comprehensive exploration of the average work done $(\langle W \rangle)$ as a function of transverse force $G$ for different values of the asymmetric parameter $b$ for a constant driving frequency $(\omega)$ and noise intensity $(D)$. The transition from an entropy-dominated region to an energy-dominated region is observed as $G$ increases.
Furthermore, an interesting trend emerges regarding the impact of the asymmetric parameter $b$ on the average work done. As the value of $b$ is incremented, there is a systematic decrease in the average work performed. This phenomenon can be explained in the following way: as the value of the asymmetric parameter $(b)$ increases, the synchronization between the time-periodic signal and particle hopping between two lobes diminishes, resulting in a decrease in the average work performed as the value of $b$ increases.
\par
\begin{figure}[!ht]
    \centering
    \includegraphics[width=0.55\textwidth]{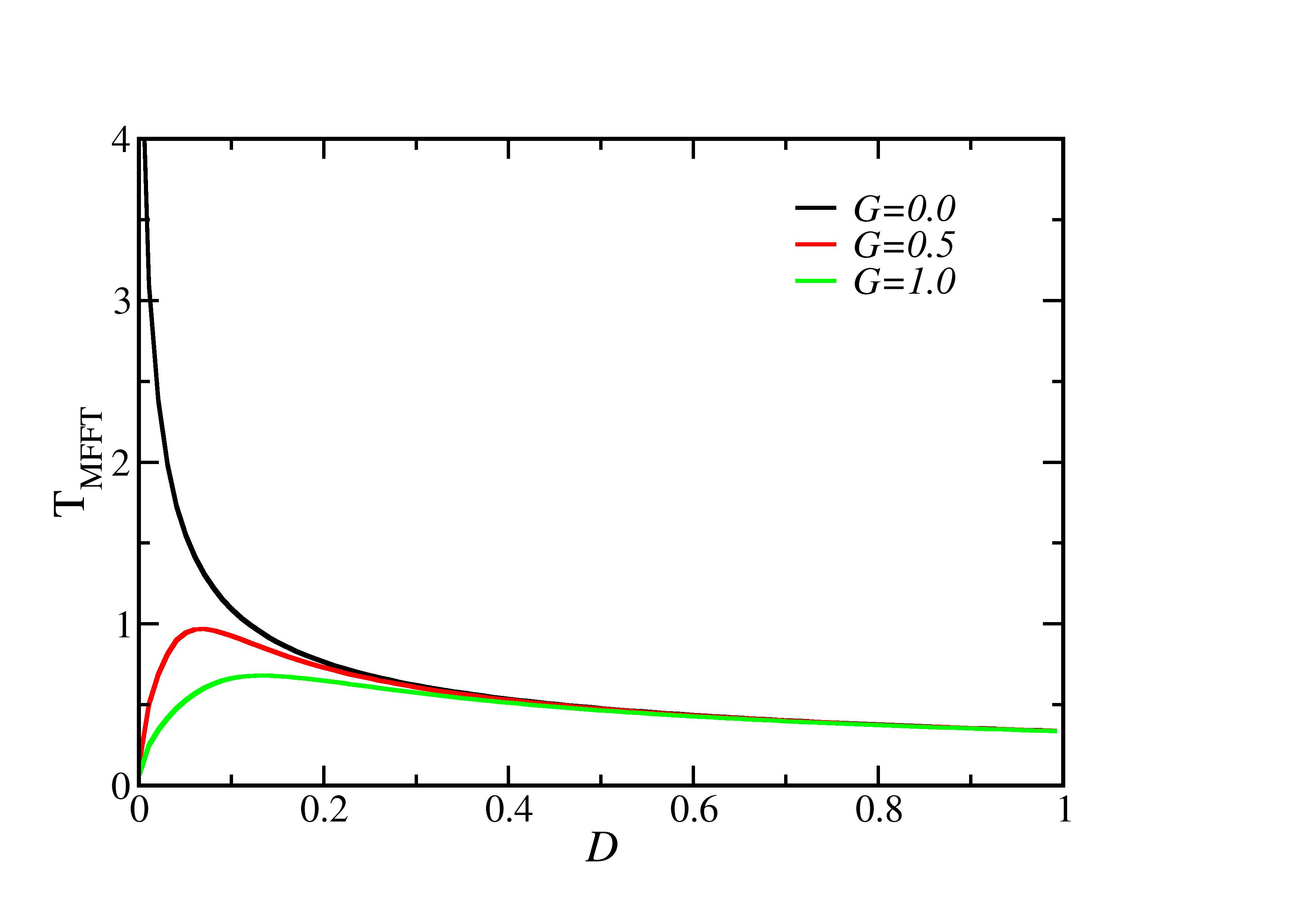}
    \caption{The variation of mean free flying time $(T_{MFFT})$ with the noise intensity ($D$) for different values of bias force $G$. For all cases $a = 0.25$, $c = 0.05$, $f_{0} = 0.03$ and $\omega=0.015$.}
    \label{f6}
\end{figure}

An alternative metric, introduced as the Mean Free Flying Time ($T_{MFFT}$), is suggested to characterize the entropic stochastic Resonance (ESR) phenomenon \cite{du2021rstc}. The instances at which the particle encounters the boundary are identified by time points $t_1$, $t_2$, $t_3$,..., $t_n$. The mean free flying time, denoted as $T_{MFFT}$, is defined as follows,
\begin{equation}\label{e7}
T_{MFFT}=\left \langle \frac{\sum_{i=1}^{n-1} t_{i+1}-t_i }{N-1} \right \rangle
\end{equation}
where the ensemble average is indicated by $\langle \rangle$. 
The entropic stochastic resonance (ESR) is noticeable in scenarios where the strength of the periodic force $(f_0)$ is small but diminishes as $(f_0)$ increases. In instances of small $(f_0)$, the synergistic effect between the longitudinal driving force $F(t)$ and noise, characterized by a moderate intensity, prevails over entropic trapping \cite{du2021rstc}.\\
The dependency among the mean free flying time $(T_{MFFT})$, noise intensity $(D)$, and the transverse bias force $(G)$ is shown in Fig.~\ref{f6}.
In the absence of the bias force $(G)$, it is intuitive that the stronger the noise intensity, the more frequently the particles collide with the wall per unit of time. In the presence of the bias force $(G)$, the competition between the entropy trapping induced by the transverse bias force $(G)$ and the escape process caused by the noise leads to an optimized mean free flying time. From the figure, we can also say that with the increasing value of $(G)$, a non-monotonic behavior of the mean free flying time only for finite values of the force $G\ne0$, while for $G=0$ the mean free flying time decays monotonically. This scenario aligns with findings in prior research \cite{Ali2024jpcb,burada_2008_entropic_prl,du2021rstc}. In other words, for the dynamics, the occurrence of the ESR effect requires a non-vanishing transverse force $(G)$. We can also conclude that the position of the mean free flying time peak shifts towards smaller noise strengths as $G$ decreases while the maximum value of $T_{MFFT}$ increases.\\

\begin{figure}[!ht]
    \centering
    \includegraphics[width=0.55\textwidth]{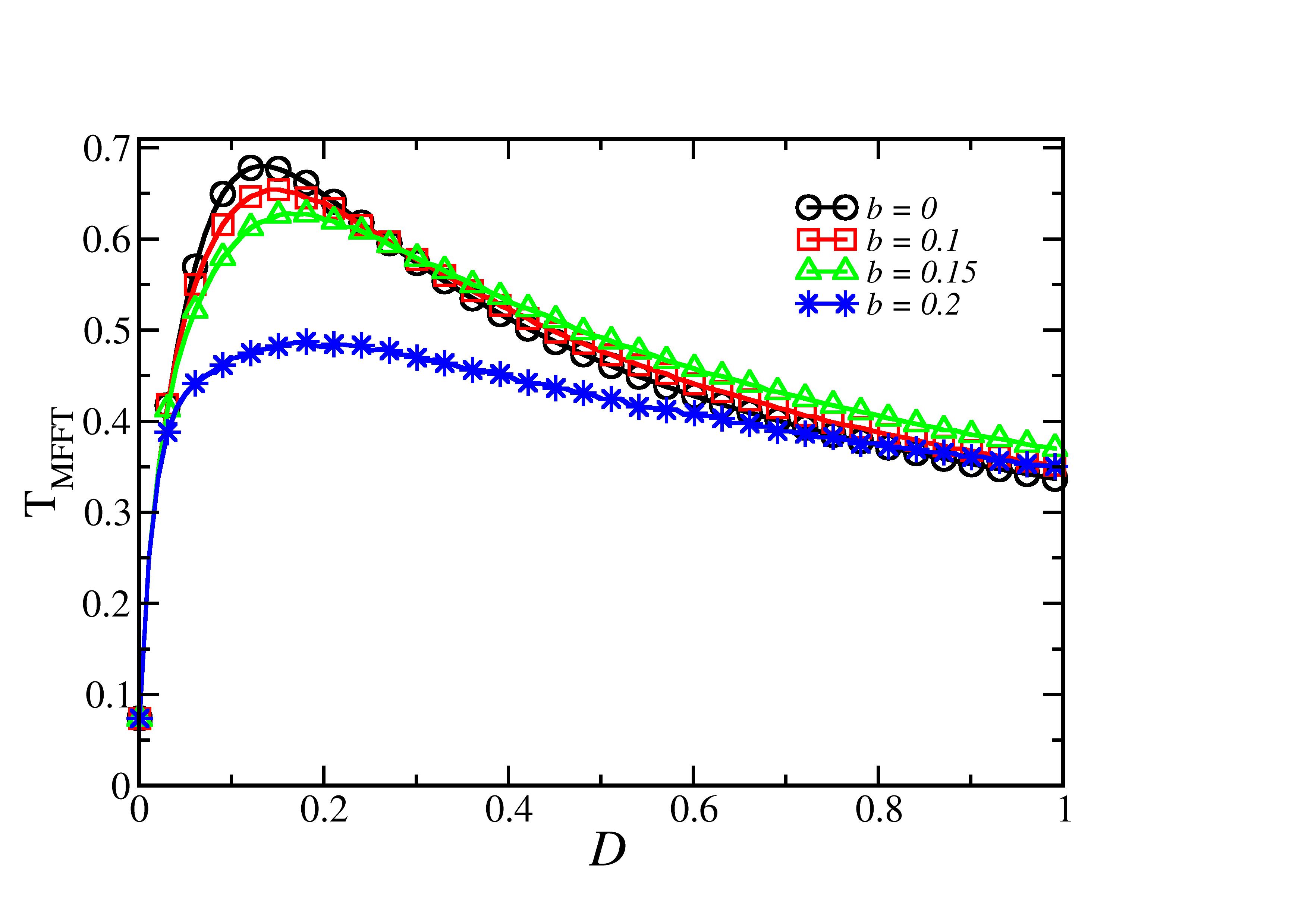}
    \caption{The variation of mean free flying time $(T_{MFFT})$ with the noise intensity ($D$) for different values asymmetric parameter $b$. For all cases $a = 0.25$, $c = 0.05$, $f_{0} = 0.03$ and $\omega=0.015$, $G=1.0$}
    \label{f7}
\end{figure}

Our main aim is to investigate the impact of asymmetry in confinement on the ESR through the measurement of mean free flying time $(T_{MFFT})$ at a finite value of transverse bias force $(G=1.0)$. To achieve this, we calculate the mean free flying time for various values of asymmetric parameters $(b)$, shown in Fig.~\ref{f7}.
The position of the $T_{MFFT}$ peak shifts towards larger noise strengths as the extent of asymmetry $(b)$ increases while the maximum value of the peak decreases.

\section{{Conclusion}\label{sec:level4}}
In conclusion, our study investigates the stochastic motion of Brownian particles in a two-dimensional asymmetrical bilobal confinement subjected to a transverse bias force and a time-periodic field. The average work done over the cycle for many periods was determined, displaying an exciting turnover. The average work changes when the drive frequency is changed. The addition of asymmetry in confinement reduced the magnitude of the maximum work performed. Furthermore, we have shown that the average work performed in the system subjected to periodic driving can be utilized as a significant metric for investigating entropic stochastic resonance.\\
\par
Additionally, we observed a transition from a state dominated by energy to one dominated by entropy as we manipulated the magnitude of the transverse force. The results of our study underscored the importance of asymmetry in confinement, as it was observed that the average work done decreases with the increase in asymmetry.\\
\par
Moreover, an alternative quantity called the mean free flying time is proposed to describe the ESR in the presence of asymmetry.\\
\par
In general, our study offers valuable insights into the intricate dynamics of particles within asymmetric confinement. It contributes to understanding the relationship between energy and entropy in stochastic systems. The discoveries enhance our comprehension of nonequilibrium phenomena and have implications across multiple disciplines, such as physics, biophysics, and nanotechnology.

\vspace{0.2cm}

\begin{acknowledgments}
The author thanks Dr. D. Mondal for his support, engaging discussions, and providing computer facilities. I also thank the Indian Institute of Technology Tirupati for providing partial financial support.
\end{acknowledgments}

\bibliography{refs}% Produces the bibliography via BibTeX.

\end{document}